\documentstyle[preprint,aps]{revtex}
\begin{document}
\title{Wealth distribution in an ancient Egyptian society}
\author{A. Y. Abul-Magd}
\address{Department of Mathematics, Faculty of Science, Zagazig University, Zagazig,\\
Egypt.}
\date{\today}
\maketitle

\begin{abstract}
Modern excavations yielded a distribution of the house areas in the ancient
Egyptian city Akhetaten, which was populated for a short period during the
14th century BC. \ Assuming that the house area is a measure of the wealth
of its inhabitants allows us to make a comparison of the wealth
distributions in ancient and modern societies.
\end{abstract}

More than a hundred years ago, Pareto \cite{pareto} proposed a power law for
the personal wealth distribution. \ He suggested that the probability
density for an individual to have a wealth (or income) of a certain value $w$
is given by 
\begin{equation}
P(w)=\frac{\alpha }{w_{\text{min}}}(w/w_{\text{min}})^{-1-\alpha }\Theta
(w-w_{\text{min}}),
\end{equation}
where $w_{\text{min}}$ is the minimal wealth, $\Theta (x)$ is the unit step
function, and the exponent $\alpha $\ is named Pareto index. \ A small value
of $\alpha $\ indicates that the individual wealth is unevenly distributed
in the corresponding society. \ The larger $\alpha $, the stronger the
suppression for larger wealths. \ Pareto analyzed the personal income for
some European countries in the 16-19th centuries. \ Some examples of the
resulting values of the Pareto index are reported in \cite{badger}. \ They
fluctuate around $\alpha =1.5$ in a wide interval ranging from $\alpha =1.13$
for Augsburg in 1526 to $\alpha =1.89$\ for Prussia in 1893. \ Analysis of
the 1935-36 US income data \cite{montroll} confirmed that the top 1\% of the
distribution follows Pareto law with $\alpha =1.63.$ \ Aoyama et al. \cite
{aoyama} clarify that the high-income range of wealth distribution of Japan
in the year 1998 follows Pareto law with $\alpha =2.06$. \ In order to
extend the analysis to the domain of intermediate and low wealths, one has
to use one of the numerous, more sophisticated models \cite{mantegna}. \ 

Recently, Solomon and collaborators \cite{solomon,biham,malcai,biham1} and
Bouchaud and M\'{e}zard \cite{bouchaud} developed an elaborate model that
describes wealth distribution, which is not restricted to large wealths. \
This model assumes that the time evolution of the income of each person is a
random multiplicative process, which can be described by a Langevin equation
with multiplicative and additive noises representing the wealths acquired
from investment and from external sources, respectively. \ In brief, the
model is given by a set of generalized Lotka-Volterra equations \cite{goel}
for the wealth $w_{i}$\ of the $i$th person 
\begin{equation}
\frac{dw_{i}}{dt}=\eta _{i}(t)w_{i}+a\overline{w}-b\overline{w}w_{i},
\end{equation}
where $\overline{w}=(1/N)\sum_{i=1}^{N}w_{i}$ is the average wealth per
capita in the society. \ The quantity $\eta _{i}(t)$ is a Gaussian random
variable of mean $m$ and variance $D$, which describes the spontaneous
growth or decrease of wealth due to various investments. \ The other terms
account for wealth redistribution due to the interaction between the members
of society and the quantities $a$ and $b$ are taken as constants in the
simplest version of the model. This equation assumes that all members of the
society exchange with each other at the same rate.\ The corresponding
Fokker-Planck equation for the wealth distribution has the following
stationary solution 
\begin{equation}
P(x)=\frac{(\alpha -1)^{\alpha }}{\Gamma (\alpha )}x^{-1-\alpha }\exp \left(
-\frac{(\alpha -1)}{x}\right) ,
\end{equation}
where $x=w/\overline{w}$ is the ratio of the personal wealth to its
expectation value and $\alpha =1+2a/D$. \ This distribution is valid for the
whole range of $w$, and has the same asymptotic behavior as Pareto
distribution. \ It has the advantage of relating the power decay of large
wealth to the wealth distribution of the poorest individual.

The purpose of the present work in to study the wealth distribution of the
society of Akhetaten (now Tell el-Amarna in Middle Egypt) in the 14th
century BC. \ This is a city founded by King Akhenaten \cite{kemp} in the
6th year of his reign that lasted for about18 years, starting from 1372 BC.
\ This king tried to replace the traditional Egyptian religion by a new
concept of god, which he called Aten. This meant that he had to destroy the
traditional pattern of religion and introduce new theology, ritual and
ecclesiastical structure. \ To begin with he changed the capital from Thebes
(now Luxor) to Akhetaten. \ Soon after King Akhenaten died, the new religion
was abandoned and the worship of the old gods was restored. \ The capital
returned to Thebes and Akhetaten was left and soon covered by sand. \
Therefore, Akhetaten was populated for 20-30 years only. It had no time to
change from generation to another as most of the other cities. In this
respect, Akhetaten is a rare example of a city, the remnants of which
reflects the state of the society which lived in it in a given time.
Moreover, its size ($\sim 1.5\times 2.0$ km)\ is typical for ancient cities
as we imagine them. One should be able to walk across any of them from one
end to another. For these reasons. Akhetaten can be considered as a fair
representative of an ancient urban society.

Modern excavations that started at the end of the 19th century and resumed
in 1977 revealed the distribution of house areas in Akhetaten \cite{kemp}.
We use this distribution to estimate Pareto index. We assume that the area
of a house $A$ is a measure of the wealth $w$ ($=x\overline{w}$) of its
inhabitants, and specifically write 
\begin{equation}
A=\overline{A}x^{\beta },
\end{equation}
where $\overline{A}$\ is the mean house area. \ With this choice, the
probability density of an area of a house in Al-Amarna excavation is
expressed in terms of the wealth distribution $P(w)$ by 
\begin{equation}
p(A)=\frac{\beta }{\overline{A}}\left( \frac{A}{\overline{A}}\right) ^{\beta
-1}P\left( \frac{A}{\overline{A}}\right) ,
\end{equation}
where $P(w)$ is given by Eq. (3). The number of houses in an interval $%
\left( A-\Delta A/2,A+\Delta /2\right) $ is given by 
\begin{equation}
N(A)=N_{0}\Delta A\frac{\beta }{\overline{A}}\frac{\left( \alpha -1\right)
^{\alpha }}{\Gamma \left( \alpha \right) }\left( \frac{A}{\overline{A}}%
\right) ^{-1-\alpha \beta }\exp \left( -\frac{\left( \alpha -1\right) }{%
\left( A/\overline{A}\right) ^{\beta }}\right) ,
\end{equation}
where $N_{0}$ is the total number of houses.

We have used Eq. (6) to calculate the area distribution of houses excavated
in Tell el-Amarna as reported by Kemp \cite{kemp}. \ The data is given as a
histogram of width $\Delta A=10$ m$^{2}$. \ The total number of houses is $%
N_{0}=$ 498. \ The mean house area is $\overline{A}=102.5\pm 3$ m$^{2}$. \
The exponents are left as free parameters. Figure shows a least-square fit
to the data of the distribution. \ The best-fit values are 
\begin{equation}
\alpha =3.76\pm 0.22,
\end{equation}
and 
\begin{equation}
\beta =0.89\pm 0.09.
\end{equation}
The resulting value of $\beta $ suggests that the area of the house is
nearly proportional to the area of the house. This seams reasonable for that
ancient society in which money was not yet invented and wages were paid in
sacks of flower and containers of beer. moreover, the houses in ancient
Egypt were built for the most part of mud brick. Thus a typical house
consisted of one floor, and this is true even for the houses of modern
Egyptian villages. Literally, the value (6) of $\beta $ suggests that seven
of every eight houses had only one floor. This is of course if we take the
total living area of the house as a measure of its value, which is true even
today.

On the other hand, the value obtained for the parameter $\alpha $ is
considerably larger than the Pareto exponents obtained for modern societies
as stated above. \ This means that the distribution of wealth in ancient
societies is narrower. \ This is expected, since ancient Egypt was a layered
society, with a veneer of bureaucracy that holds most of the total wealth on
the top of a vast underlayer of peasants and craftsmen.\ The middle class
exists practically only after the industrial revolution. \ 

We can deduce more information about the ancient society under consideration
from the observed value of the exponent $\alpha $ by using the following
arguments, which have been raised by Solomon and Richmond\cite{richmond}. \
The distribution (3) decays to zero extremely fast as one goes to lower\
values of $x$ below its maximum. \ In this respect, it resembles the simpler
distribution (1), which can be used to define an effective $w_{\text{min}}$
in terms of the mean wealth $\overline{w}$ by 
\begin{equation}
x_{\text{min}}\equiv \overline{w}/w_{\text{min}}=1-1/\alpha .
\end{equation}
Equation (7) suggests that $w_{\text{min}}=073\overline{w},$\ which means
that most of the populations were living near the poverty line. \ In
addition to that, let $L$\ be the average number of dependents supported by
an average wealth. The poorest people, who cannot even afford a family, will
ensure that they do not earn less than the share of a member of an average
family. \ The minimum wealth $w_{\text{min}}$ is\ then estimated as $%
\overline{w}/L$. Thus, according to (6) 
\begin{equation}
L=\alpha /(\alpha -1).
\end{equation}
Accordingly, the average number of dependents in Akhetaten $L=1.36\pm 0.25$,
which means that in an average family, two persons of every three had to
work. In other words, children had to work in a very early age. \ 

We would like, however, to note that the results of the present paper are
based on the assumption that Eq. (2) is a realistic stochastic model of
wealth distribution. The value of $\alpha $ obtained above strongly depends
on the relationship between the income of the rich and poor, as determined
by Eq. (3) which is based on the Lotka-Volterra model. \ Indeed, the house
distribution \ resulting from the excavation under consideration fluctuates
considerably for areas $A$ \ larger that twice the mean value $\overline{A}$%
. \ Moreover, the data cover only the domain of $A<4\overline{A}$. \ Thus
the distribution does not extend far enough in the asymptotic region. To
show the effect on the evaluation of Pareto index, we replaced the function $%
\exp \left[ -(\alpha -1)/x\right] $ in (3) by another function that reaches
unity faster, e.g., $\exp \left( -b/x^{2}\right) $, so that 
\begin{equation}
P(x)=\frac{2b^{\alpha /2}}{\Gamma (\alpha /2)}x^{-1-\alpha }\exp \left( -%
\frac{b}{x^{2}}\right) ,
\end{equation}
with $b=\left[ \Gamma (b/2)/\Gamma (b/2-1/2)\right] ^{2}$. \ We applied this
formula to the analysis of the wealth distribution in Aketaten. The best-fit
vale of the exponent that we obtained for the resulting distribution is $%
\alpha =1.59\pm 0.19$, which agrees very well with the values of Pareto
index obtained for contemporary societies. \ However, the quality of fit
will not be as good for values of $x$ lower than the position of the peak.

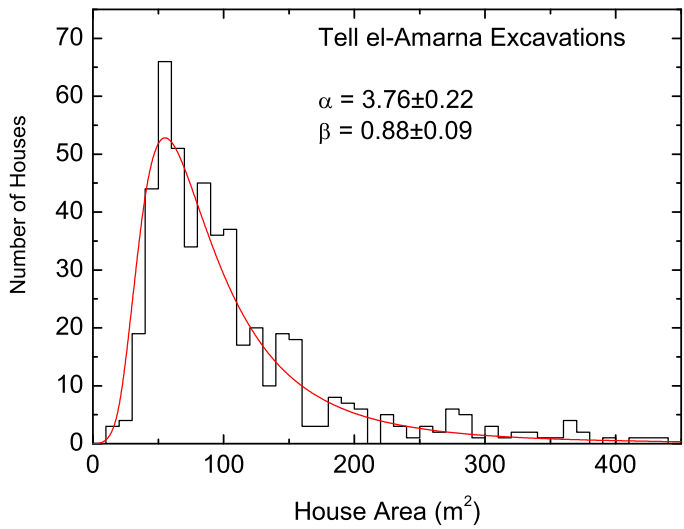
\end{document}